\documentstyle[11pt,aaspp4]{article}

\def\aa{A\&A}

\def\ergs{\,ergs s$^{-1}$ }

\def\>{$>$}
\def\<{$<$}
\def\sun{$_{\odot}$ }
\def\arcmin{$^{\prime}$ }
\def\nh{N$_{\rm H}$ }

\def\deg{$^{\circ}$~ }
\def\gx{GX 354-0~}
\def\ks{KS 1731-260~}

\received{}
\accepted{}

\begin{document}

\title{ASCA Observations of GX 354-0 and KS 1731-260}

\author{T. Narita\altaffilmark{1}, J.E. Grindlay}
\affil{Harvard-Smithsonian Center for Astrophysics, Cambridge, MA 02138, USA}
\and
\author{D. Barret}
\affil{Centre d'Etude Spatiale des Rayonnements, Toulouse Cedex 04, France}


\altaffiltext{1}{tnarita@cfa.harvard.edu} 


\begin{abstract}
We report on ASCA observations of the low mass X-ray binaries GX 354-0
and KS 1731-260. The spectrum of \gx is best described as a power-law
or a Comptonized spectrum with $\tau\sim5$ and kT$\sim8$ keV and a
residual at $\sim6.5$ keV. The residual may be a disk reflection or a
Compton broadened Gaussian line from the hot inner ADAF-like coronal
region. The absorption column density to the source is
$2.9\times10^{22}$ cm$^{-2}$. No soft thermal component was
detected. The spectrum from \ks is softer and it is best fit with a
two component model with a column density of $1.1\times10^{22}$
cm$^{-2}$. The likely interpretation is emission from a Comptonizing
cloud with an optical depth $\tau>12$ and either a neutron star or a
disk blackbody emission.  We discuss the likely location of the
Comptonizing cloud for both sources within the context of several
proposed emission models.

\end{abstract}



\section{Introduction}

The X-ray emission process from low-mass X-ray binaries (LMXBs) is
thought to depend primarily on the accretion flow structure and its
interaction with the compact object. Early observations showed that
several high luminosity sources could be fit with a two component
blackbody and disk blackbody model (\cite{mitsuda84},
\cite{white88}). However, low luminosity sources could be described
with either a single component model, such as a power-law or thermal
Bremsstrahlung (see White, Nagase \& Parmar 1993 and
references therein), or two
component models involving a harder component and a softer blackbody
(\cite{callanan95}).  In an  
attempt to describe the emission process, White et al. (1985)
described that the thermal Bremsstrahlung-like component in the
spectrum of Sco X-1 was due to an unsaturated Comptonization of
accretion disk photons in the corona. Previously, a saturated
Comptonization model had also been used to explain the hard X-ray
emission from Cyg X-1 (\cite{sunyaev80}).  Recent observations of
LMXBs with the broad-band detectors on Rossi X-ray Timing Explorer
(RXTE) and BeppoSAX have confirmed that in addition to a hard
Comptonized emission, an additional soft thermal component is indeed
present in many low luminosity LMXBs
(\cite{guainazzi98}, \cite{barret99}, \cite{piraino99},
\cite{barret00}, \cite{bloser00}).  It now seems likely that many 
X-ray binaries possess both a hard and a soft component in their
spectrum regardless of the luminosity.  In this paper, we continue the
study of the spectra of low luminosity LMXBs by examining two X-ray
bursters, \gx and KS 1731-260.  Our intent was to use the higher
spectral resolution of ASCA to measure the absorption
column to the sources, and to constrain the spectral shape so that we
may further study the origin of the emission source.

\gx (also MXB 1728-34) and \ks are Atoll LMXBs
(\cite{hasinger89}) located in the
galactic bulge.  Of the two, \gx is the better studied, particularly
since its discovery as a burst source (\cite{lewin76}, 
\cite{hoffman76}).  Previous observations of \gx with Einstein,
EXOSAT, Ginga, and ROSAT have shown a spectrum that can be fit with
either a thermal Bremsstrahlung of kT $\sim 13-18$ keV, or a power-law
of photon index $\sim1.7-2.1$ (\cite{grindlay81}, \cite{white86}, \cite{day90},
\cite{foster86}, \cite{schulz99}). The column density is large but
uncertain, ranging from 1.5 to 3.5$\times10^{22}$ cm$^{-2}$
(\cite{basinska84}, \cite{schulz99}). Most recent observations with
RXTE and BeppoSAX have found a spectrum best fit with a Comptonization
model with a soft component, and a 6.7 keV iron line (\cite{disalvo00},
\cite{piraino00}). 

\ks was discovered as a transient burster
(\cite{sunyaev89}), but it has since been observed as a persistent
source by GINGA, ROSAT, SIGMA, and RXTE (\cite{yamauchi90},
\cite{predehl95}, \cite{barret92}, \cite{barret00}).  The Mir-Kvant/TTM
observation of \ks found a 5.7 keV 
thermal Bremsstrahlung spectrum absorbed by $2.2\times10^{22}$
cm$^{-2}$ of neutral hydrogen (\cite{sunyaev89}).  A subsequent ROSAT
observation found a power law fit with an \nh of
$1.3\times10^{22}$ cm$^{-2}$ (\cite{barret98}). Recent observations by
RXTE found the spectrum to be best described by a Comptonized
component and a softer thermal component (\cite{barret00}). 

The
distance estimates to \gx range from 4--14 kpc (\cite{grindlay81}),
although 
\cite{paradijs78} estimated a distance to the source of $4.2\pm0.2$
kpc assuming the maximum X-ray burst flux was Eddington limited for
1.4 M\sun object.  RXTE observed a burst with photospheric radius
expansion from KS 1731-260, from which Smith et al. (1997) derived a distance
of $8.3\pm0.3$ kpc assuming a 1.4 M\sun neutron star.  In this paper,
we use the distances of 4.3 kpc and 8.3 kpc for \gx and KS
1731-260. Both sources, being near the Galactic center, are heavily
reddened and no optical counterparts have been found.  The possible
reddened globular cluster identification for \gx reported by Grindlay
and Hertz (1981) was not confirmed by more sensitive imaging IR
observations (\cite{isaacman89}).

\section{Observation and results}

\gx and \ks were observed by ASCA on September 27, 1997, with both
the SIS and GIS detectors. The total observing time for \gx was $\sim9$
ksec.  Nearly half of the observing time was telemetered in the medium and
low rates which were saturated due to the relatively high count rate.
In this analysis we used the high telemetry mode data to derive the
flux estimate, but we combined the high and medium telemetry mode data
for the spectral fits.  The observing time for KS 1731-260 was $\sim8.5$
ksec, and all the data were recorded in the high telemetry mode. No
other sources were detected in either observations' field of view.

Initially, we attempted to make the GIS and SIS data consistent. While
the GIS residuals from fitting the Crab spectra are only few percent
at 2 keV, there is a growing divergence between the GIS and the SIS
spectra below 1 keV (\cite{ascahelp}). The quantum efficiency of the
SIS detectors has been slowly degrading due to radiation damage, and
the current detector calibrations do not account for the loss of
efficiency.  When we compared the GIS and SIS
standard screened data sets (Rev 2) using the latest redistribution
matrix and ancillary response files, we found that the flux differed
by as much as 15\% between the two detectors at energies between 1 to
2 keV and at energies greater than 8 keV.  By selecting only the
single pixel events (grade 0) in the SIS, we were able to reconcile
the high energy discrepancy between the GIS and SIS-1. However, the
response of SIS-0 over-predicts the model when only grade 0 events are
used (\cite{ascahelp}), and thus we chose to use only the SIS-1 data
for further analysis.  For the low energy difference, we first tried
various extraction diameters for the source, and various extraction
regions for the background.  Since the SIS-1 local background count
rate was $<1\%$ of the source count rate for \gx and KS 1731-260, this
did not affect the spectral shape.  Second, we looked for pileup in
the SIS-1 data by examining the difference in the spectrum inside and
outside the 0.5\arcmin source radius.  The difference in the best fit
spectral parameters for \gx and \ks was within $2\sigma$, which
indicated that the pileup was not affecting the SIS-1 data. Finally,
we used strict screening criteria in {\tt ascascreen}.  Observations
only at angles $>40$\deg from the bright Earth limb and COR $<6$ GeV
were accepted. However, this also did not modify the low energy
spectrum.  Other authors have also noted the loss in low energy
response from the SIS appearing as an increase in the inferred column
density (Hwang et al. 1999, \cite{mukai99}).  We found that an
additional SIS-1 absorption of $4.0\times10^{21}$ cm$^{-2}$ for GX
354-0, and $2.4\times10^{21}$ cm$^{-2}$ for KS 1731-260 was required
in our observation.  To avoid the large systematic error from the
uncertain SIS calibration at low energy, we used only the SIS-1 data
from 2.0 to 10.0 keV in further joint analysis with the GIS.


We used a standard screening criteria to extract the GIS events.  The
data were rejected during the time when the Earth limb elevation angle
was $<5^{\circ}$ and during regions of low geomagnetic rigidity ($<4$
GeV).  The data were also spatially screened for background and
calibration events near the edge, and non X-ray events, as determined
from their rise times, were rejected.


\subsection{GIS Light curves}

We extracted the GIS light curves for both sources and examined them
for any temporal variability in the flux. \ks did not show any
variability exceeding $2 \sigma$.  There was one type-I X-ray burst
detected from \gx but the variability in the remainder of the light
curve did not exceed $2 \sigma$.  The All Sky Monitor (ASM)
aboard RXTE showed the flux from GX 354-0 and KS 1731-260 were
$\sim90$ mCrab and $\sim140$ mCrab respectively in the 1.3--12 keV
energy 
band.  While the ASM count rate from KS 1731-260 was near the maximum
level during the time of our observation, the count rate from GX 354-0
remained at a constant level before and after our observation.

Since the ASM has three energy bands, we plotted the hardness ratio
(5.0-12.0/3.0-5.0 keV) for both sources from the time ASM began monitoring
each source (Fig.~\ref{fig1-asm}).  We found the mean hardness ratio
of \gx was larger than that of KS 1731-260, indicating that \gx
generally has a harder spectrum. 

From the ASCA data, we also constructed a color-color diagram for the
persistent emission from both sources (Fig.~\ref{fig2-cc}).  We used
the GIS-3 data and designated 1.0-3.0 keV for the soft band, 3.0-5.0
keV for the medium band, and 5.0-10.0 keV for the hard band. We did
not find strong spectral variability in either observation. Since
our observation was short, we cannot comment on the exact position
within an Atoll source color-color diagram.

\subsection{\gx X-ray burst}

At 23:09:13 UT, a moderate sized type-I X-ray burst was detected from
GX 354-0.  The satellite was in the high telemetry mode, and the peak
count rate in the GIS was $\sim100$ cts sec$^{-1}$. The burst showed an
exponential decay with an e-folding time $15\pm3.5$ seconds. In the
persistent emission spectral analysis, we excluded the burst data from
23:08:40 to 23:11:20 UT.  The deadtime corrected burst data was binned
into 6 consecutive time intervals (labled A--F in
Figure~\ref{fig3-burstlc}), and spectral analysis was done on each
interval using the combined GIS-2 and GIS-3 detectors
(Fig.~\ref{fig4-burst}).  The persistent emission was used as the
background, and the spectra were fit using a simple blackbody model
(BB) with neutral absorption. The BB temperature and normalization
were allowed to vary, but the column density was fixed at the value
derived from the persistent emission fits. The best-fit parameters are
listed in Table~\ref{burst_tab}. The BB temperature peaked at 
$\sim2.1$ keV, and the normalization increased asymptotically. No evidence
of photospheric expansion was seen. If a 4.3 kpc distance to the source
is assumed, the BB radius increased from 1 km to $\sim10$ km in 
roughly 10 seconds and remained at $\sim10$ km radius for the next
20 seconds. The peak flux was $1.0\times10^{-8}$\ergs (1--10 keV),
which corresponds to a luminosity of $2\times10^{37}$ ergs s$^{-1}$ at
4.3 kpc.

\subsection{Persistent emission spectral analysis}

\subsubsection{\gx}

The spectrum from \gx was analyzed by simultaneously fitting the high
and medium telemetry mode data for the GIS-2, GIS-3, and SIS-1
detectors.  The GIS photon extraction region was limited to the
standard 6\arcmin radius centered on the source, and the background
was taken from an annulus with an inner diameter of 6\arcmin and an
outer diameter of 12$^{\prime}$. The energy range was restricted to
0.8-10.0 keV band.  After deadtime correction, the effective
observation time was 6370 seconds. The high 
telemetry mode source count rates after background subtraction were
$23.03\pm0.07$ and $28.12\pm0.08$ s$^{-1}$ for the GIS-2 and GIS-3 
detectors. The spectrum was binned using {\tt grppha} with a minimum
of 400 counts per bin. The SIS source extraction region was 3.26\arcmin
and the background was combined from various regions in the detector
field of view. The energy range was restricted to 2.0-10.0 keV. The
effective observation time was 7582 seconds, and the background
subtracted source count rate was $11.46\pm0.06$ s$^{-1}$ for the high
telemetry mode.
The spectrum was binned with a minimum of 100 counts per bin.
The spectral fits were done using XSPEC v10.0.

We first tried to fit the simultaneous three detector data with a
simple absorbed power law or a Comptonization model (CompST). This 
resulted in a power law photon index of $\alpha=1.98$ with a
$\chi^{2}_{\nu}=1.23$ (1036 dof), or a
Comptonizing cloud with an optical depth $\tau=7.8$ and electron
kT$=5.9$ keV with a $\chi^{2}_{\nu}=1.22$ (1035 dof). The GIS
absorption column 
was $2.90\times10^{22}$ cm$^{-2}$ for both models. In the residuals,
there was an excess at $\sim6-7$ 
keV. This excess could not be adequately fit with a high temperature
blackbody such as the one found in 4U 0614+091 (\cite{piraino99}).
Adding a Gaussian line ($\chi^{2}_{\nu}=1.05$) or a disk
reflection component ($\chi^{2}_{\nu}=1.10$) improved the fit. The
best fit Gaussian line was broad ($\sigma\sim 1.29\pm0.30$ keV) with a
line centroid of 6.4 keV. The equivalent width of the line was 550 eV. 
Figure 5 shows the power 
law fit residuals with and without the Gaussian component.  For the disk
reflection model 
(PEXRAV), the inclination was fixed to 60 degrees and the abundances
were fixed at solar.
The reflection
fraction, which is proportional to the solid angle subtended by the
disk at the illuminating source, was allowed to vary.
The best fit gave a reflection fraction of $\sim2$. 
To see whether a weak soft component was also present, we added a soft
blackbody to both the pow+Gaussian and pow+Refl models. The $\chi^2$
improvement was only marginal ($\Delta\chi^2<1$).  The standard F-test
showed that the addition of the blackbody component was not
significant at the 95\% confidence level ($F=0.18$, $\Delta\nu=2$).
We list the best fit spectral parameters in
Table~\ref{gxfits_tab}. The total 1--10 keV flux was
$2.5\times10^{-9}$ ergs cm$^{-2}$ s$^{-1}$.

Since the SIS detector has better energy resolution than the GIS, we
examined the SIS-1 data alone in the 0.5-10 keV band for additional weak
lines.  We looked in particular for line
features in the residual near 0.7 keV due to ionized Fe or O
emissions and the narrow Gaussian line at 1.6 keV reported by Di Salvo
et al. (2000).  The additional Gaussian components in the model, however,
did not improve the fit from the continuum model, and we set an
equivalent width upper limit of 17 eV for a $\sigma=0.05$ keV line
at 1.6 keV. 

\subsubsection{KS 1731-260}

We used a similar 6\arcmin radius GIS source extraction region and a
3.26\arcmin SIS-1 source region.  The energy band was restricted to
0.8-10 keV band for the GIS and 2.0-10 keV band for the SIS-1.  The data
were deadtime corrected, and the background was also extracted in a
similar manner to GX 354-0.  The background subtracted source count
rate in the GIS-2 and GIS-3 detectors were $37.05\pm0.07$ and
$43.27\pm0.07$ s$^{-1}$, while the grade 0 SIS-1 source count rate was
$14.76\pm0.04$ s$^{-1}$. The total observation time was 7620 seconds
and 8772 seconds for the GIS and SIS respectively. The GIS data were
also regrouped into spectral bins with a minimum of 1000 counts per
bin and 100 counts per bin for the SIS-1.

The simultaneous three detector fit found a much softer spectrum
for KS 1731-260.  The spectrum was best fit with a combined
Comptonization model with a soft thermal component ($\chi^{2}_{\nu}=1.08,
\nu=677$).
The two component fit (CompST+bb) is shown in Figure 6.
From the combined Comptonization and a soft component model, 
we could not uniquely identify the soft emission as a simple blackbody
or a disk blackbody. However, in the CompST+disk model, the ratio of
the disk blackbody luminosity to the total luminosity was $\sim0.5$,
while in the CompST+bb model the blackbody luminosity fraction was
only 0.17.  A single component Comptonization model
($\chi^{2}_{\nu}=1.37, \nu=679$) or power law model
($\chi^{2}_{\nu}=2.29, \nu=680$) was excluded.  A two component model
involving a power law and a blackbody was also excluded
($\chi^{2}_{\nu}=1.33, \nu=678$).  
We list the spectral
parameters in Table~\ref{ksfits_tab}. 
The total 1-10 keV band flux was $3.20\times10^{-9}$ ergs
cm$^{-2}$ s$^{-1}$, of which the blackbody flux was
$0.55\times10^{-9}$ ergs cm$^{-2}$ s$^{-1}$ and the disk blackbody
flux was $1.73\times10^{-9}$ ergs cm$^{-2}$ s$^{-1}$.

Barret et al. (2000) found a broad iron line in their analysis of RXTE
data.  Our fits did not require a line, and 
in Table 3 we report the equivalent width upper limits for a
similar ($\sigma=0.8$ keV) 6.4 keV iron line (90\%
confidence).  We also did not find any narrow line features at
$\sim0.7$ keV in the SIS-1 data. 

\section{Discussion}

We observed \gx and \ks in the 1-10 keV energy band and found them to
be in very different spectral states. The spectrum from \gx was best
fit with a hard power law ($\alpha\sim2$) or a hot Comptonization
model (kT$\sim8-9$ keV) with either a broad ($\sigma\sim1.3$ keV)
weakly ionized iron line emission or a disk reflection (reflection
fraction $\sim2.0$).  A soft thermal component was not seen from GX
354-0. The spectrum from \ks was softer and was best described as a
combination of an optically thick Comptonization component and a
second blackbody component.  The comparison of the derived 1-10 keV flux to
other pointed observations (Table~\ref{flux_tab}) showed that ASCA
observed \gx in a relatively 
low flux state and \ks in a middle to high flux state.  
The corresponding 1-10 keV luminosities of the sources at an assumed
distances of 4.3 kpc and 8.3 kpc are $5.2\times10^{36}$ and
$2.5\times10^{37}$ ergs sec$^{-1}$, which are consistent with the ASM
count rate on the day of observation.

With the advances in detector sensitivity, relatively low luminosity
LMXBs Atoll sources
have been shown to have both a Comptonized emission and a soft thermal
emission, like their high luminosity counterparts. The question
remains however whether the soft and the Comptonized emissions
originate in the same regions regardless of the accretion rate
(\cite{barret00}, \cite{bloser00}). Mitsuda et al. (1989) found that a
multi-color disk and a Comptonizing boundary layer could explain the
emission during a high state of 4U 1608-522, but a number of authors
have pointed out that the soft thermal component seen in the low state
of other Atoll sources could come from either the disk or from an
optically thick boundary layer (e.g. \cite{white88},
\cite{guainazzi98}, \cite{barret99}, \cite{piraino99}).

If the Comptonized emission originates in the compact boundary layer
and the soft emission comes from the disk, the inferred inner disk
radius in \ks ($\sim10$ km) is in good agreement with a canonical
neutron star radius.  The inner disk temperature of 0.9 keV is also
consistent with that observed from 4U 1608-522 and 4U 1820-30 in their
high accretion states (\cite{mitsuda89}, \cite{bloser00}).  For GX
354-0, the lack of soft emission could indicate that the disk
truncates at a large radius.
If the excess at 6-7 keV is interpreted as an iron fluorescence line,
the line 
energy of 6.4 keV again suggests that the emission is due to the
fluorescent emission from the neutral iron in the cool outer disk and
not from the hot inner disk. 

Alternatively, the Comptonized emission could arise in the extended
accretion disk corona (ADC) while the soft component originates near
the neutron star surface.  For KS 1731-260, the blackbody temperature
(0.6 keV) is comparable to the blackbody temperatures found in similar
LMXBs (\cite{callanan95}, \cite{guainazzi98}, \cite{schulz99},
\cite{barret99}, \cite{piraino99}), but the best fit blackbody 
radius for \ks (24 km) is large for a neutron star.  In the case of GX
354-0, an extended ADC subtending a large angle to the disk could
result in either a disk reflection component with a large reflection
fraction, or a broadened line emission.  The best fit value of the
reflection fraction is somewhat uncertain since we did not have the
high energy coverage to constrain the reflection hump. However, the
large disk reflection fraction is similar to the values seen from GS
1826-34 and 4U 1608-522 in their low hard states (\cite{zdziarski99}).
If the excess at $6-7$ keV is instead due to the fluorescence emission
from the disk, Sunyaev and Titarchuk (1980) found that 
photons escaping from a Comptonizing cloud will result in an estimated
line broadening of $\Delta\lambda\sim\lambda_{c}\tau^2$, where
$\lambda_{c}$ is the Compton wavelength and $\tau$ is the optical
depth.  From the observed line width $\sigma\sim1.3$ keV we derive an
optical depth $\tau\sim4.3$, consistent with the best fit value from
GX 354-0.

Our results from \ks are best described with a compact
Comptonizing boundary layer, and the results from \gx are seemingly
consistent 
with either of the emission models.  However, both models suffer from
other observational and theoretical difficulties.  In the case of a
Comptonizing boundary layer, the hard emission is confined to be near
the neutron star surface, $r\sim10^{6}$ cm for 1.4M\sun neutron star.
Yet, in the analysis of dipping LMXBs, the light curve is best
described by a partial covering of an extended corona ($>10^{9}$ cm)
and a compact soft source (\cite{church98}). A compact hard emission
region would also seem unlikely to produce the broad emission line or
the large disk reflection seen from GX 354-0.  In the case of an extended
Comptonizing corona with an optically thick boundary layer, Sunyaev
\& Shakura (1986) showed the ratio of the luminosity from the boundary
layer should be equal or larger than the luminosity from the corona,
if the disk extends to the marginally stable orbit. Yet, the
luminosity ratio of the boundary layer to the corona is typically too
small (\cite{white88}, \cite{guainazzi98}, \cite{barret99},  
\cite{piraino99}). Our observation is no exception; the luminosity
ratio of the boundary layer to the corona is only 0.24 for \ks and
even less for GX 354-0.  The energy release in the boundary layer may
be reduced if the neutron star spins at near breakup (\cite{white88}),
but neither of the neutron stars (\gx at 363 Hz, \cite{strohmayer98};
\ks at 260 Hz, \cite{wijnands97}) appear to rotate at speeds close to
break up. 

Another possible emission mechanism is that the boundary layer may be
optically thin during periods of low accretion
(\cite{barret00}). Citing spectral similarities between neutron star
LMXBs and black hole LMXBs in their low states, Barret et al. (2000)
suggests that an optically thin accretion flow such as an
advection-dominated solution of Narayan \& Yi (1995) could replace the
standard inner disk in LMXBs.  If an optically thin accretion flow
could transition to the neutron star surface in an optically thin
boundary layer, the spectrum would be dominated by Comptonized
emission, and only a small fraction of the flux would come from the
neutron star surface or the accretion disk (\cite{barret00}).

In such a scenario, it is plausible that we have observed the
interaction of the cold outer disk with the Comptonizing corona in the
low accretion state of GX 354-0, namely the 6.4 keV iron line or the
reflection from a weakly ionized disk.  The broadening of the iron
line may be due to the dispersion in the corona if the corona subtends
a large solid angle as seen from the disk.  At higher accretion rate,
the optically thin corona is expected to collapse to an optically
thick disk (Narayan \& Yi 1995). Correlation of the high frequency
quasi-periodic oscillation to the spectral parameters also gives good
indication that the disk moves toward the central source during
periods of high accretion (\cite{kaaret99}, \cite{mendez99},
\cite{bloser00}). One would then expect to observe more emission from
the hot ionized disk at higher accretion rate.  This is consistent
with the 6.7 keV iron line and the $\sim1$ keV disk blackbody
component reported 
by Di Salvo et al. (2000) and Piraino et al. (2000) in their
observations of \gx during higher flux states.  
Similarly, a soft component, in addition to the Comptonizing
component, was also 
observed from KS 1731-260 in its medium and high 
flux states with ASCA and RXTE.  Our observations with ASCA did not
find the broad 6.4 keV line that was seen by RXTE, however the upper
limit equivalent width set by our ASCA observations
is marginally consistent with that found by RXTE.  
Since the energy
resolution of the PCA detector on board RXTE is 18\% at 6 keV, it
is possible that the the broad line seen with RXTE was a superposition
of ionized and non-ionized lines.  Presumably, such a state would
exist during a transitions between the low and high accretion states,
where the emissions from both the outer disk and the inner disk are
visible. 
Further observations using the high resolution gratings on
Chandra or XMM may provide more direct evidence on the strength of
the line emissions at various accretion rates. Such study should be
valuable in constraining the extent of the ADC and its interaction
with the accretion disk.

\acknowledgments

The ASM quick look results were provided by the ASM/RXTE teams at MIT
and at the RXTE SOF and GOF at NASA's GSFC.  This work was supported
in part by NASA grants NAG5-5103 and NAG5-5209.

\clearpage

\begin{deluxetable}{crrrrrr}
\footnotesize
\tablecaption{GX 354-0 X-ray Burst Spectral Parameters. \label{burst_tab}}
\tablewidth{0pt}
\tablehead{
\colhead{Phase\tablenotemark{a}} & \colhead{T$_{\rm
exp}$\tablenotemark{b} (sec)} & 
\colhead{CR (sec$^{-1}$)} & \colhead{kT$_{\rm bb}$ (keV)} & 
\colhead{R$_{\rm bb}$ (km)} & \colhead{F$_{\rm x}$\tablenotemark{c}} &
\colhead{$\chi^2$/dof}
}
\startdata

A & 2.44 & $28.0\pm4.5$ & $1.81\pm0.40$ & $3.04\pm0.97$ & 0.45 &
17.7/21 \nl  
B & 2.85 & $63.2\pm5.4$ & $2.11\pm0.28$ & $3.72\pm0.65$ & 1.09 &
46.2/45 \nl 
C & 3.46 & $55.1\pm4.7$ & $1.53\pm0.11$ & $5.42\pm0.66$ & 0.71 &
35.9/49 \nl
D & 3.10 & $62.0\pm5.4$ & $1.15\pm0.07$ &$ 9.39\pm1.16$ & 0.55 &
38.1/47 \nl 
E & 3.65 & $45.0\pm4.5$ & $1.08\pm0.07$ & $9.48\pm1.26$ & 0.45 &
45.4/47 \nl 
F & 11.05 & $32.7\pm2.3$ & $0.89\pm0.04$ & $10.64\pm1.07$ & 0.23 &
100.5/113 \nl 
\enddata

\tablenotetext{a}{Time interval in burst (cf. Fig. 3)} 
\tablenotetext{b}{Deadtime corrected exposure time} 
\tablenotetext{c}{1-10 keV flux, $10^{-8}$ erg cm$^{-2}$ s$^{-1}$ } 

\tablecomments{\nh is fixed at $3.1\times10^{22}$ cm$^{-2}$. Errors
are quoted at 90\% confidence level for one interesting parameter.} 

\end{deluxetable}

\clearpage

\begin{deluxetable}{lcccccc}
\footnotesize
\tablecaption{\gx Persistent Emission Spectral Parameters. \label{gxfits_tab}}
\tablewidth{0pt}
\tablehead{ 
\colhead{} &
\colhead{Pow} &
\colhead{Pow+ga} &
\colhead{Pow+refl} &
\colhead{CompST} &
\colhead{CompST+ga} &
\colhead{CompST+refl}
} 
\startdata

\nh\tablenotemark{a} & $2.90_{-0.03}^{+0.03}$ & $3.09_{-0.05}^{+0.07}$
& $3.13_{-0.04}^{+0.06}$ & $2.89_{-0.02}^{+0.02}$ &
$3.09_{-0.06}^{+0.07}$ & $3.16_{-0.05}^{+0.05}$ \nl

$\alpha$ & $1.98_{-0.02}^{+0.01}$ & $2.15_{-0.05}^{+0.06}$ &
$2.24_{-0.05}^{+0.03}$ & & & \nl

kT & & & & $5.99_{-1.22}^{+41.02}$ & $8.93_{-3.39}^{+58.73}$ &
$8.32_{-3.16}^{+23.57}$ \nl
 
$\tau$ & & & & $7.84_{-6.42}^{+1.05}$ & $5.69_{-4.47}^{+1.76}$ &
$5.55_{-1.07}^{+1.74}$ \nl


E & & $6.35_{-0.14}^{+0.12}$ & & & $6.36_{-0.14}^{+0.12}$ & \nl

$\sigma$ & & $1.29_{-0.24}^{+0.33}$ & & & $1.29_{-0.24}^{+0.33}$ & \nl

Refl\tablenotemark{b} & & & $2.05_{-1.27}^{+7.95}$  & & & $1.85_{-0.92}^{+8.15}$ \nl

F$_{tot}$\tablenotemark{c} & 2.53 & 2.51 & 2.53 & 2.52 & 2.50 & 2.53\nl


$\chi^{2}_{\nu}$ (dof) & 1.23 (1036) & 1.05 (1033) & 1.10 (1035) &
1.22 (1035) & 1.05 (1032) & 1.10 (1032) \nl

\enddata

\tablenotetext{a}{Absorption column, $10^{22}$ cm$^{-2}$}
\tablenotetext{b}{Reflection fraction hard limit set at 10}
\tablenotetext{c}{Total flux 1.0--10.0 keV, $10^{-9}$ erg cm$^{-2}$ s$^{-1}$}
\tablecomments{Errors are quoted at 90\% confidence level for one
interesting parameter.} 

\end{deluxetable}

\clearpage

\begin{deluxetable}{lccccc}
\footnotesize
\tablecaption{\ks Persistent Emission Spectral Parameters. \label{ksfits_tab}}
\tablewidth{0pt}
\tablehead{\colhead{} & \colhead{CompST} & \colhead{CompST+bb} &
\colhead{CompST+dbb} &
\colhead{Pow+bb} & \colhead{Pow+dbb}
} 
\startdata

\nh\tablenotemark{a} & $1.44^{+0.01}_{-0.02}$ & $1.06^{+0.07}_{-0.07}$
& $1.08^{+0.01}_{-0.02}$ & $1.37^{+0.03}_{-0.03}$ &
$1.30^{+0.03}_{-0.03}$ \nl

kT & $2.24^{+0.06}_{-0.08}$ &
$1.93^{+0.09}_{-0.09}$ & $1.82^{+0.06}_{-0.07}$ & & \nl


$\tau$  & $12.71^{+0.28}_{-0.30}$ & $18.37^{+2.17}_{-1.72}$ &
$40.59^{+\infty}_{-9.17}$ & & \nl

$\alpha$ & & & & $2.15^{+0.03}_{-0.03}$ & $2.05^{+0.05}_{-0.06}$ \nl

kT$_{bb}$  & & $0.58^{+0.02}_{-0.01}$ & $0.93^{+0.01}_{-0.01}$ &
$0.85^{+0.03}_{-0.03}$ & $1.28^{+0.06}_{-0.06}$ \nl

R$_{bb}$\tablenotemark{c}  & & $23.5^{+2.3}_{-2.6}$ & &
$8.69^{+0.88}_{-0.40}$ & \nl

R$_{in}\sqrt{cos\theta}$\tablenotemark{c}  & & & $11.9^{+0.6}_{-0.5}$
& & $4.27^{+0.52}_{-0.46}$ \nl

F$_{tot}$\tablenotemark{d}  & 3.20 & 3.20 & 3.20 & 3.20 & 3.20 \nl

F$_{bb}$\tablenotemark{e} & & 0.55 & 1.73 & 0.59 & 1.22 \nl

$\chi^{2}_{\nu}$ (dof)  & 1.37 (679) & 1.08 (677) & 1.09 (677) &
1.33 (678) & 1.27 (678) \nl

EW\tablenotemark{f} & $<45.7$ & $<65.4$ & $<118.0$ & & \nl
\enddata

\tablenotetext{a}{Absorption column, $10^{22}$ cm$^{-2}$} 
\tablenotetext{b}{Assumed distance of 8.3 kpc}
\tablenotetext{c}{Total flux 1.0--10.0 keV, $10^{-9}$ erg cm$^{-2}$ s$^{-1}$}
\tablenotetext{d}{Blackbody flux 1.0--10.0 keV, $10^{-9}$ erg cm$^{-2}$ s$^{-1}$}
\tablenotetext{e}{Equivalent width (eV) upper limit (90\% confidence)
for a 6.4 keV line ($\sigma=0.8$ keV)
}
\tablecomments{Errors are quoted at 90\% confidence level for one
interesting parameter.} 

\end{deluxetable}

\begin{deluxetable}{lccccc}
\footnotesize
\tablecaption{Comparison of Flux with Previously Reported Fits
\label{flux_tab}} 
\tablewidth{0pt}
\tablehead{ \colhead{Detector} & \colhead{Model} &
\colhead{Flux\tablenotemark{a}} &
\colhead{Line Energy\tablenotemark{b}} & \colhead{Line
Width\tablenotemark{c}} & \colhead{Refs}}  
\startdata
\multicolumn{6}{c}{GX 354-0}\\ 
\em{BeppoSAX} MECS & Comp+bb & 4.4 & 6.7 & 0.34 & 1\\
\em{BeppoSAX} MECS & Comp+bb & 3.1 & 6.7 & 0.5 & 2\\
\em{Einstein} MPC & TB & 2.9 & & & 3\\
\em{ASCA} GIS+SIS & Comp & 2.5 & 6.4 & 1.3 & 4\\
\hline
\multicolumn{6}{c}{KS 1731-260} \\
\em{RXTE} PCA & Comp+bb & 8.1 & 6.4 & 0.8 & 5\\
\em{ASCA} GIS+SIS & Comp+bb & 3.2 & & & 4\\
\em{Kvant} TTM & TB & 0.22 & & & 6\\
\enddata

\tablenotetext{a}{1.0--10.0 keV, $10^{-9}$ erg cm$^{-2}$ s$^{-1}$} 
\tablenotetext{b}{line centroid, keV} 
\tablenotetext{c}{$1\sigma$ width, keV} 

\tablerefs{(1) \cite{piraino00}, (2) \cite{disalvo00},
(3) \cite{grindlay81}, (4) This paper, (5) \cite{barret00}, (6)
\cite{sunyaev90}}  
\end{deluxetable}

\clearpage

\clearpage

\figcaption[ASMlcurve.eps]{RXTE/ASM hardness ratio (5--12 keV/
3--5 keV) of \gx and \ks from MJD 50083 (1/1/96). The observation date
is marked with an arrow.  The dotted line shows the weighted mean of
the hardness ratio since ASM began monitoring each source. The
spectrum from \ks is typically softer than the spectrum from
GX 354-0. The spectrum from \ks may be particularly soft on the day of our
observation.  \label{fig1-asm}}

\figcaption[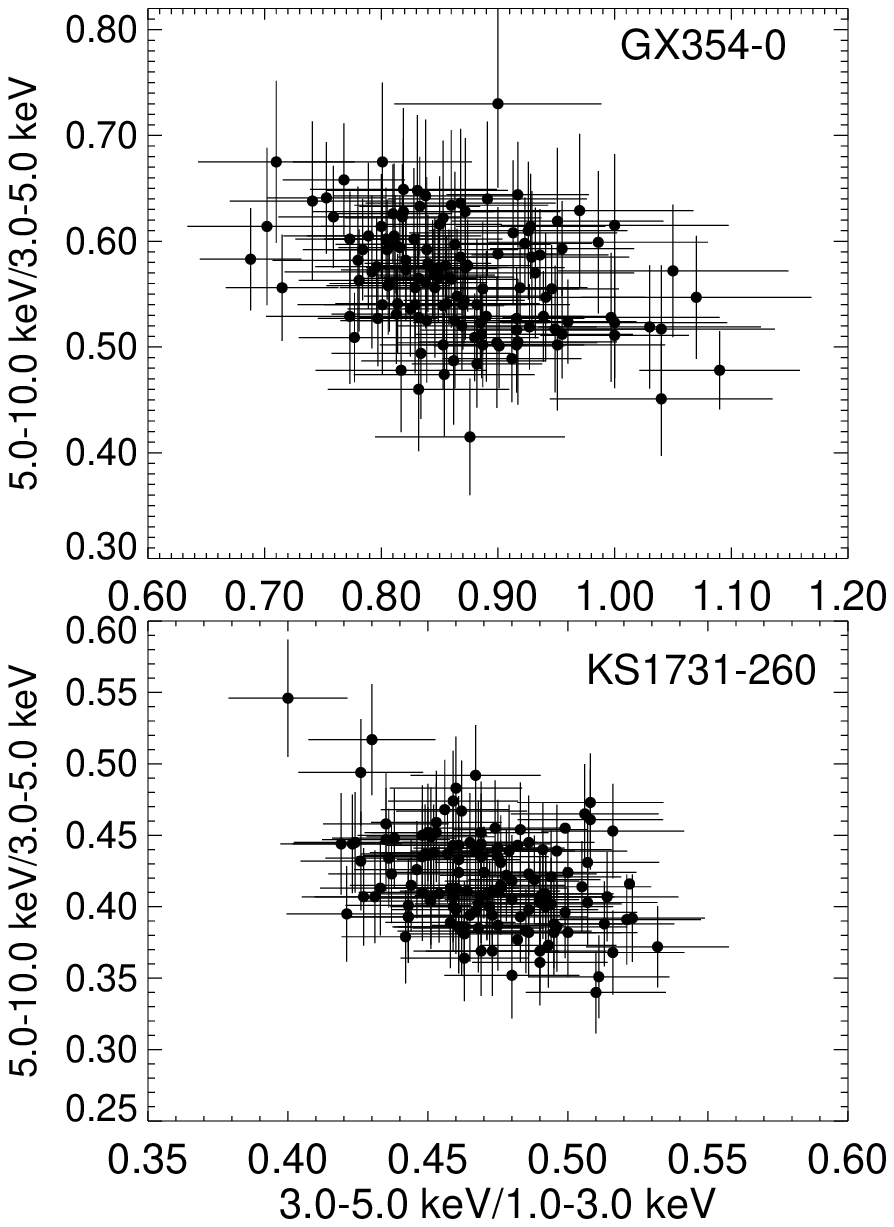]{The color-color diagrams for \gx and \ks
computed from the GIS-3 data. The high and medium telemetry data on
\gx are combined. Neither source showed spectral variations during the
pointed observation.  \label{fig2-cc}}

\figcaption[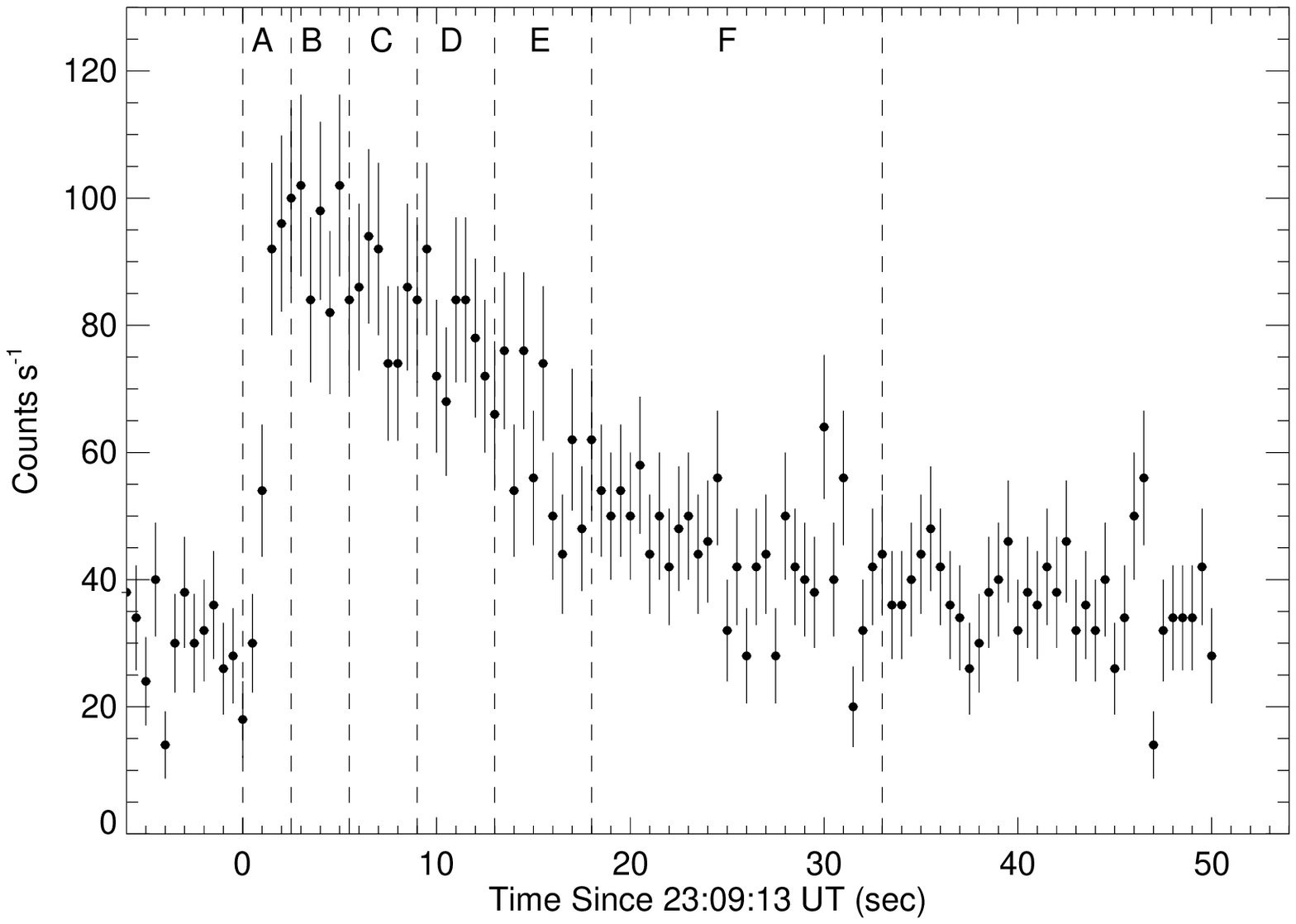]{A type I X-ray burst light curve from
\gx seen by the GIS-3 detector.  The data was binned in 0.5 second
intervals, and the burst was subdivided into 6 time resolved segments
(A--F) for spectral analysis.  \label{fig3-burstlc}}

\figcaption[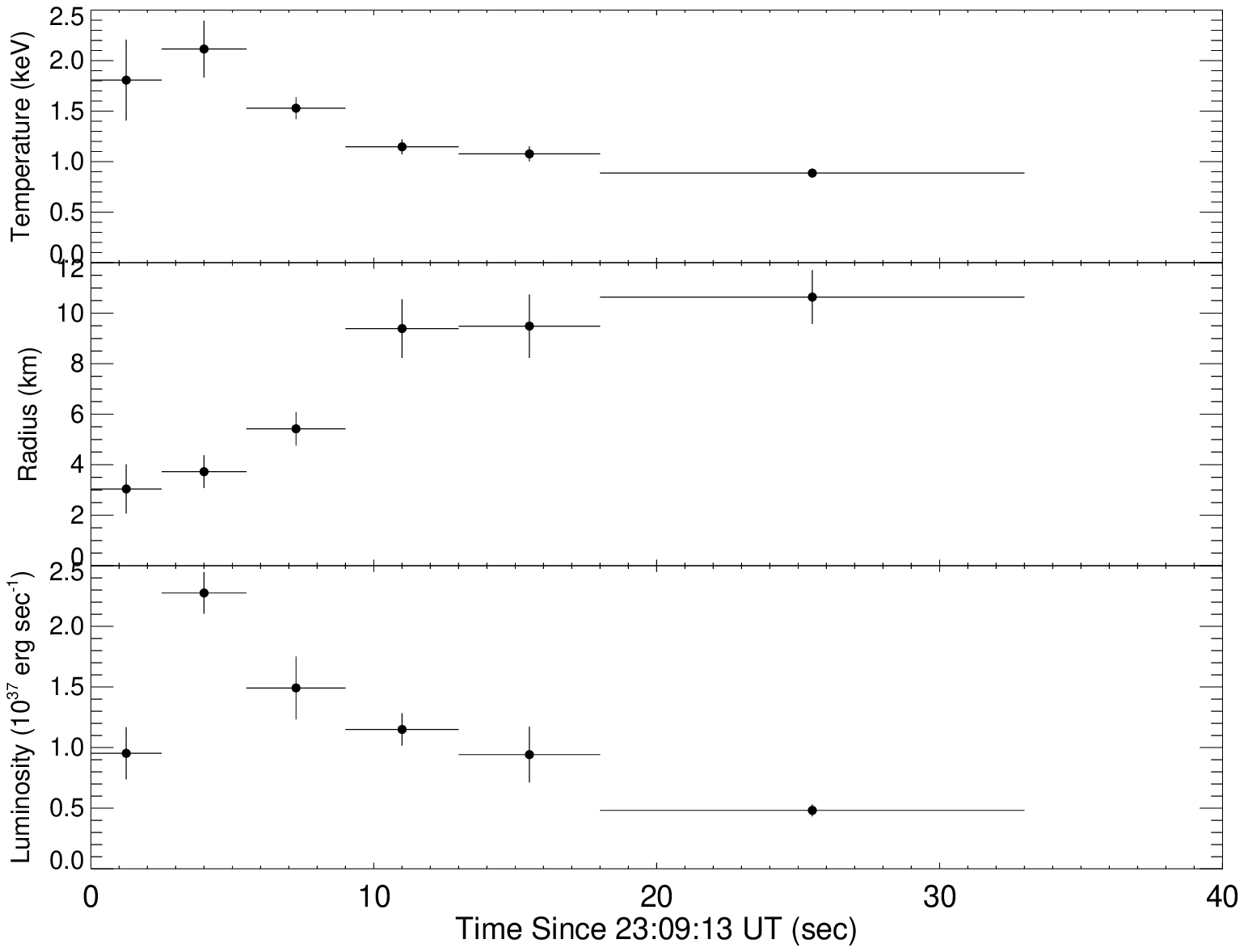]{The time evolution of the spectral
parameters during the X-ray burst. The parameters were derived from
simultaneous fits of the GIS-2 and GIS-3 detectors. The absorption
column density was fixed at $3\times10^{22}$ cm$^{-2}$, the best fit
absorption from the persistent emission analysis. \label{fig4-burst}}

\figcaption[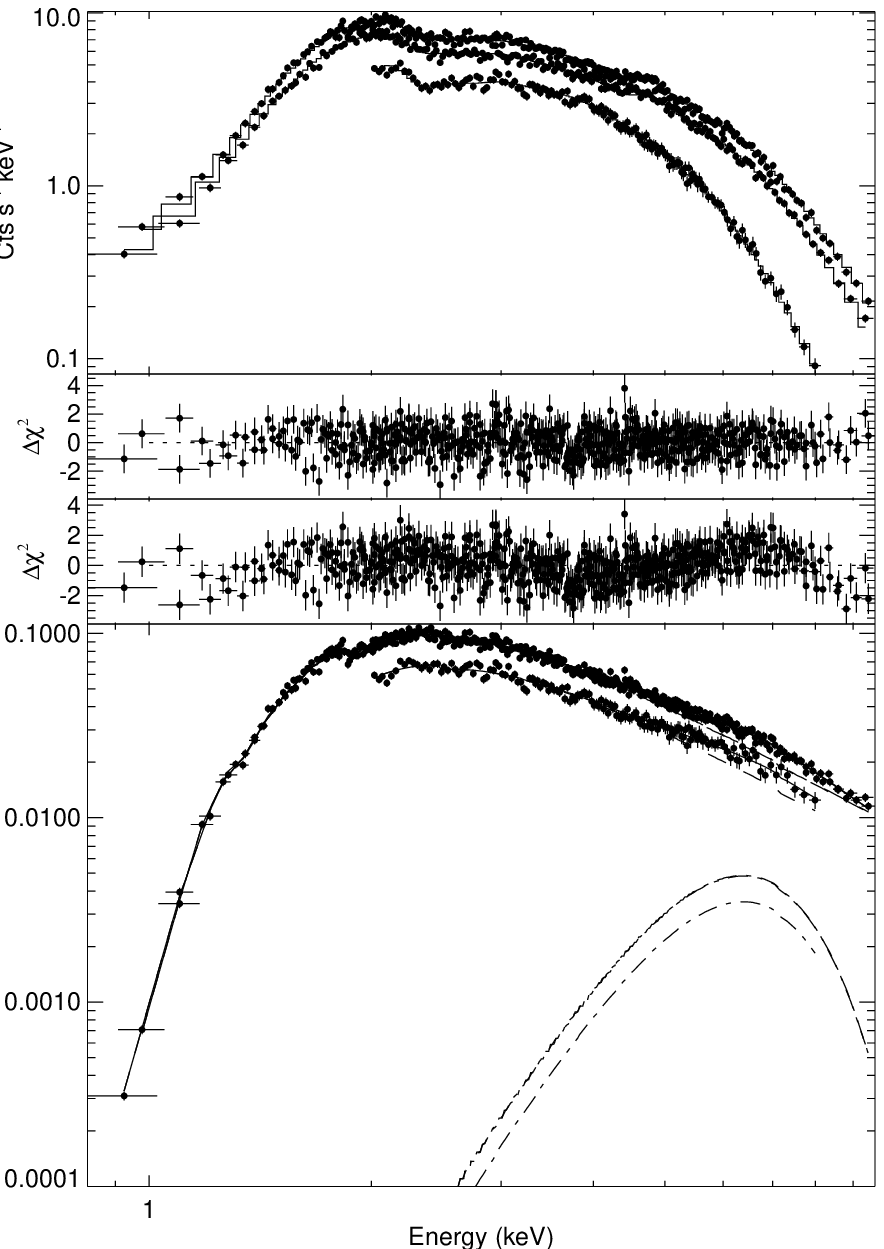]{The GIS and SIS-1 spectra of \gx
simultaneously fit with the power-law + Gaussian model.  For clarity,
only the high telemetry data is shown. The top panel
shows the raw count rate spectrum with the fitted model folded through
the instruments' responses, the second panel shows the residuals of
the fit, the third panel shows the residual without the additional
Gaussian component, and the bottom panel shows the unfolded spectrum
with the individual model components.  The dot-dashed line is the
Gaussian, and the dashed line is the power-law model. \label{fig5-gx}}

\figcaption[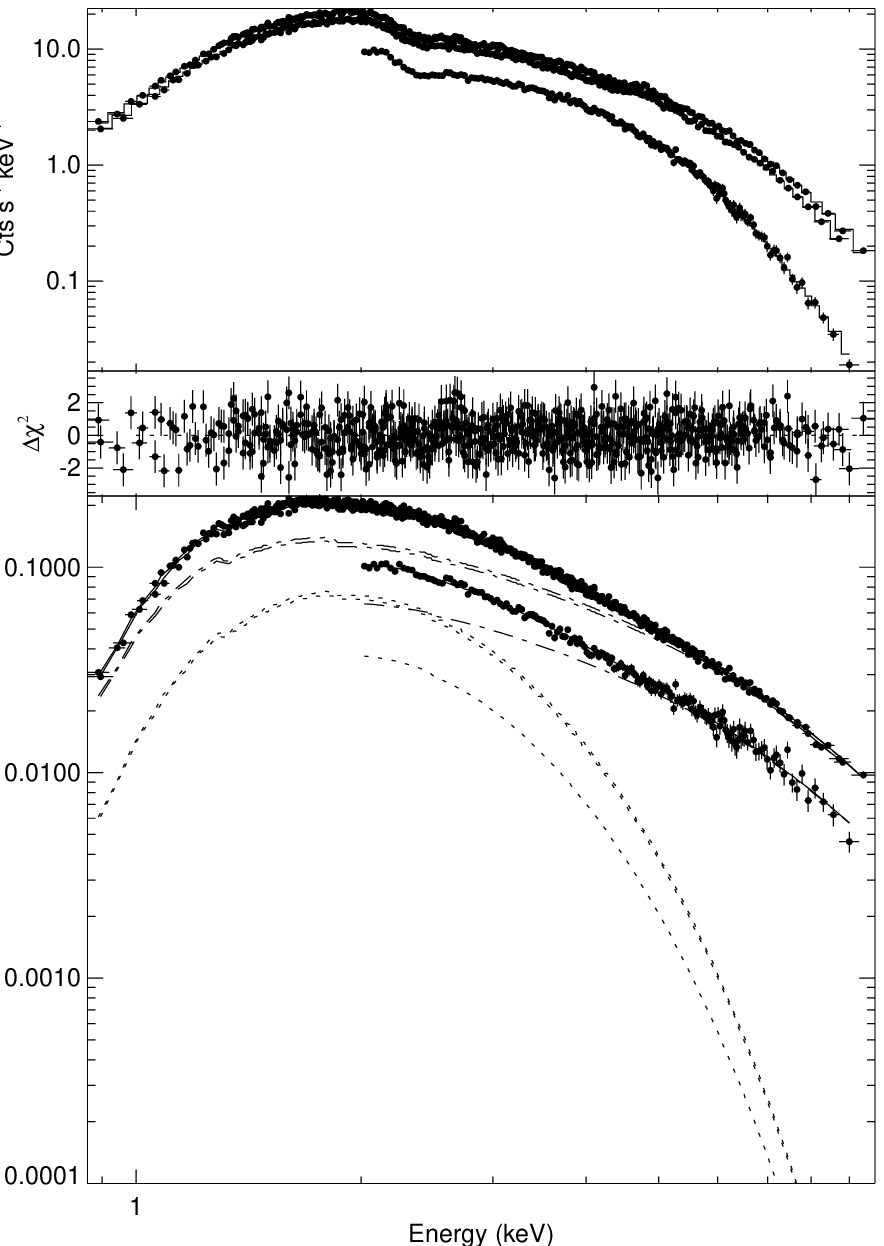]{The GIS and SIS-1 spectra of \ks
simultaneously fit with the CompST + BB model. The top panel shows the
raw count rate spectrum with the fitted model folded through the
instruments' responses, the middle panel shows the residuals of the
fit, and the bottom panel shows the unfolded spectrum with the
individual model components. The dotted line is the blackbody, and the
dot-dashed line is the CompST model. \label{fig6-ks}}



\clearpage

\plotone{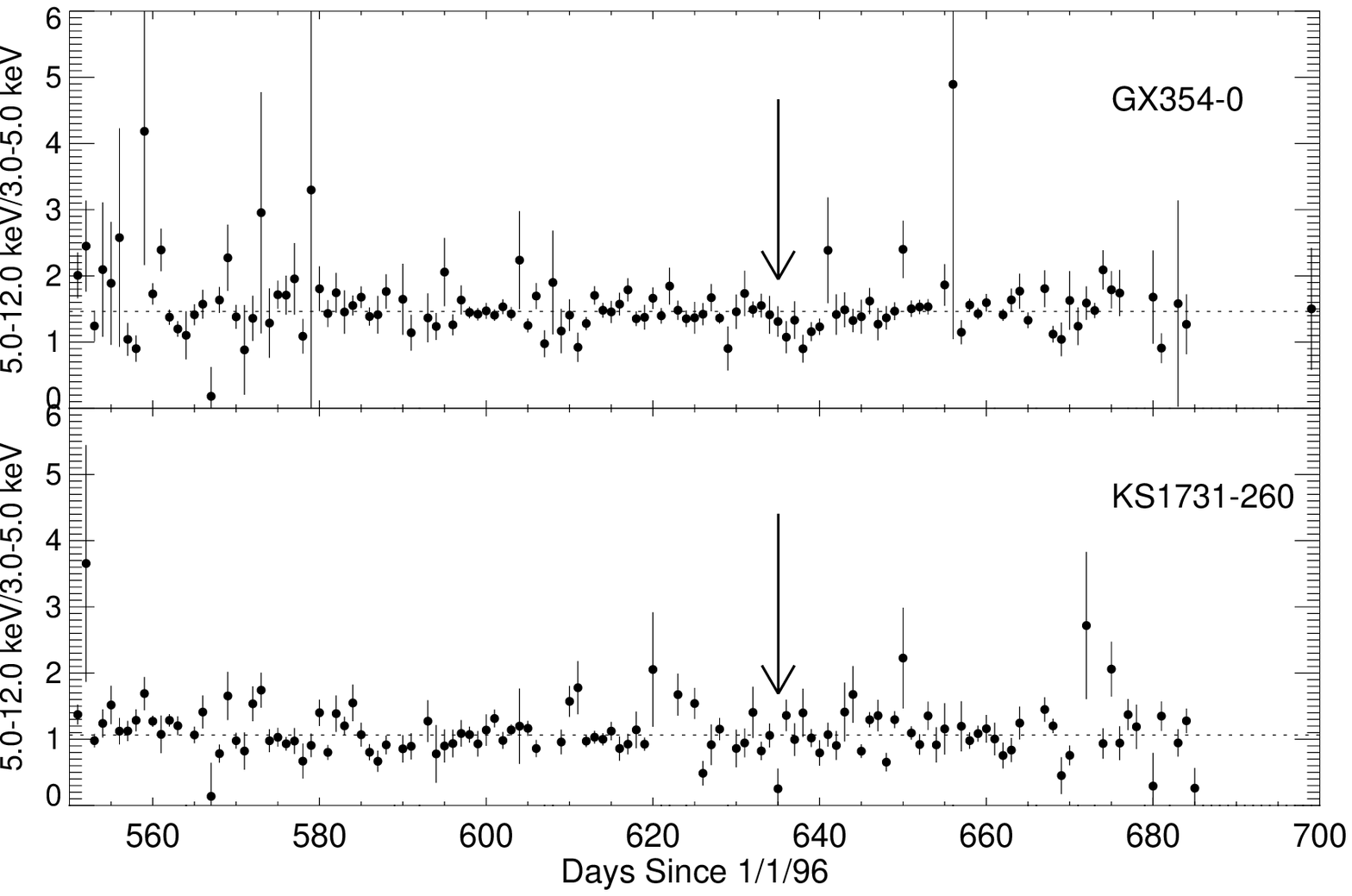}

\clearpage

\plotone{GIS3cc.eps}

\clearpage

\plotone{burstlc.eps}

\clearpage

\plotone{burstparams.eps}

\clearpage

\plotone{gxfit_spec.eps}

\clearpage

\plotone{ksfit_spec.eps}

\end{document}